\documentclass[aps,prl,twocolumn,groupedaddress]{revtex4-1}

\usepackage{color}

\usepackage[utf8]{inputenc}
\usepackage[english]{babel}
\usepackage{amsmath,graphicx,enumerate}

\begin{document}


\title{Spontaneous velocity alignment in Motility-induced Phase Separation} 

\author{L. Caprini$^{1}$}
\author{U. Marini Bettolo Marconi$^{2}$}
\author{A. Puglisi$^{3}$} 
\affiliation{$^1$ Gran Sasso Science Institute (GSSI), Via. F. Crispi 7, 67100 L'Aquila, Italy.\\ $^2$
Scuola di Scienze e Tecnologie, Universit\`a di Camerino - via Madonna delle Carceri, 62032, Camerino, Italy.\\ $^3$ Istituto dei
  Sistemi Complessi - CNR and Dipartimento di Fisica, Universit\`a di
  Roma Sapienza, P.le Aldo Moro 2, 00185, Rome, Italy }

\date{\today}


\begin{abstract}


We study a system of purely repulsive spherical 
self-propelled particles in the minimal set-up inducing Motility-Induced 
Phase Separation (MIPS). We show that, even if explicit alignment 
interactions are absent, a growing order in the velocities of the 
clustered particles accompanies MIPS.  
Particles arrange into aligned or vortex-like domains. Their sizes increase as the persistence of the self-propulsion grows, an effect that is quantified studying the spatial correlation function of the velocities.
We explain the velocity-alignment by unveiling a hidden alignment interaction of the Vicsek-like form, induced by the interplay between steric interactions and self-propulsion.  
As a consequence, we argue that the MIPS transition cannot be fully understood in terms of a scalar field, the density, since the collective orientation of the velocities should be included in effective coarse-grained descriptions.

\end{abstract}

\maketitle


Fishes~\cite{ward2008quorum}, birds~\cite{ballerini2008interaction} or
insects~\cite{attanasi2014finite} often display fashinating collective
behaviors such as flocking~\cite{ballerini2008interaction,
  mora2016local} and swarming~\cite{cavagna2017dynamic}, where all
units of a group move coherently producing intriguing dynamical
patterns.  A different mode of organization of living organisms is
clustering, for instance in bacterial colonies~\cite{dell2018growing},
such as E. Coli~\cite{berg2008coli}, Myxococcus xanthus~\cite{peruani2012collective} or Thiovulum majus~\cite{petroff2015fast}, relevant for histological cultures in several
areas of medical and pharmaceutical sciences.  Out of the biological
realm, the occurrence of stable clusters~\cite{bialke2015active,
  palacci2013living, buttinoni2013dynamical, ginot2018aggregation},
stable chains~\cite{yan2016reconfiguring} or vortices~\cite{bricard2015emergent} in activated colloidal particles,
e.g. autophoretic colloids or Janus disks~\cite{howse2007self, takatori2016acoustic},
offers an interesting challenge for the design of new materials.

Even if the microscopic details differ case by case, a few classes of
minimal models with common coarse-grained features have been
introduced in statistical physics. Units in these models are called
``active'' or ``self-propelled'' particles~\cite{marchetti2013hydrodynamics, ramaswamy2010mechanics,
  bechinger2016active} to differentiate them from Brownian colloids
which passively obey the forces of the surrounding environment.
Propelling forces may be either of mechanical origin (flagella or body
deformation), or of thermodynamic nature (diffusiophoresis and
self-electrophoresis)~\cite{palacci2010sedimentation,
  theurkauff2012dynamic}.  In some simple and effective examples,
self-propulsion is modeled as a constant force with stochastic
orientation, as in the case of Active Brownian Particles (ABP)~\cite{ten2011brownian, romanczuk2012active}.  Thermal fluctuations
  play only a marginal role and stochasticity is usually due to the
  unsteady nature of the swimming force itself.

It is well-known that dumbells, rods and, in general, elongated microswimmers display a marked orientational order even in the absence of alignment interactions \cite{peruani2006nonequilibrium, aranson2003model, ginelli2010large, deseigne2010collective}.
Instead, in the literature, it is believed that explicit aligning velocity-interactions are crucial to observe velocity alignment between spherical self-propelled units~\cite{vicsek2012collective}. 
This kind of interaction, such as that in the seminal Vicsek model~\cite{vicsek1995novel}, consists in a short-range force that aligns the velocity of a target particle to the average of the neighboring ones. 
Vicsek interactions lead to
long-range polar order~\cite{toner1995long, toner2012reanalysis, mahault2018active}, density
inhomogeneities in the form of traveling bands~\cite{gregoire2004onset, solon2015phase} or periodic density waves~\cite{caussin2014emergent}. 
Recently, models with orientation-velocity couplings have been implemented to obtain a global polar order without assuming any explicit velocity-alignment between neighboring particles \cite{lam2015self, giavazzi2018flocking}.
Instead, the interplay
between steric interactions and self-propulsions is recognized to be
the minimal requirement for phase-separation in self-propelled systems.
This occurs even in the absence of any attractive
force~\cite{gonnella2015motility}, at variance with passive Brownian
particles.  Such a phenomenon, known as Motility-induced Phase
Separation (MIPS) has been largely investigated~\cite{cates2015motility}, starting from the pioneering work of Fily
and Marchetti~\cite{fily2012athermal}.  
The coexistence of clustering
and velocity ordering has been recently considered, 
and, even if its role in MIPS is still an open question~\cite{sese2018velocity, barre2015motility, van2019interrupted, shi2018self}, 
it has been shown that may induce freezing in dense regimes~\cite{geyer2019freezing}.
The alignment, characterizing Vicsek-like models~\cite{chate2008modeling}, and the ABP
phase-separation are phenomena which 
are usually thought to be generated  by two distinct types
of interactions between particles.

In the present study, we challenge the widespread idea that explicit
alignment interactions are necessary to  observe a growing orientational order or
- equivalently - that 
 the velocity alignment
observed in Vicsek-like models
do not appear in purely repulsive, spherical ABP particles.  To the best of our
knowledge, previous studies aimed to measure the polarization, i.e.
the existence of a common orientation of the self-propelling force,
but overlooked the possibility of ordering in the real particles'
velocity, that is the crucial observation of the present report.
 
We consider a suspension of $N$ interacting self-propelled particles,
for simplicity (and without loss of generality) in two dimensions.
The evolution of the center of mass coordinate of each microswimmer,
$\mathbf{x}_i$, is described by an over-damped equation of motion with
self-propulsion embodied by a time-dependent external force with
constant modulus, $v_0$, and orientation vector, $\mathbf{n}_i$, of
components $(\cos{\theta_i}, \sin{\theta_i})$. According to the ABP
scheme, the orientational angles, $\theta_i$, evolve as independent
Wiener processes. Interactions are purely repulsive and no explicit
aligning forces are included. Therefore the dynamics reads:
\begin{subequations}
\label{eq:wholeABPdynamics}
\begin{align}
\label{eq:xf_dynamics}
\gamma\dot{\mathbf{x}}_i &= \mathbf{F}_i + \gamma v_0 \mathbf{n}_i  \\
\label{eq:theta_dynamics}
\dot{\theta}_i&= \sqrt{2D_r} \xi_i  \,,
\end{align}
\end{subequations}
being $D_r$ the rotational diffusivity (thermal diffusion is usually
negligible) while $\gamma$ is the constant drag coefficient.  Steric
interactions are modeled by the force $\mathbf{F}_i=-\nabla_i U_{tot}$,
being $U_{tot} = \sum_{i<j} U(|{\mathbf r}_{ij}|)$ with ${\mathbf r}_{ij}= \mathbf{x}_i -\mathbf{x}_j$. We choose
$U(r)$, with as a purely repulsive potential of the WCA type, namely
$U(r)=4\epsilon\left[\left(\frac{\sigma}{r}\right)^{12}-
  \left(\frac{\sigma}{r}\right)^{6}\right] + \epsilon$, for $r\leq
2^{1/6}\sigma$ and zero otherwise.  The constant $\sigma$ represents
the nominal particle diameter while $\epsilon$ is the energy scale due
to interactions. 



\begin{figure}[!t]
\centering
\includegraphics[width=0.9\linewidth,keepaspectratio]{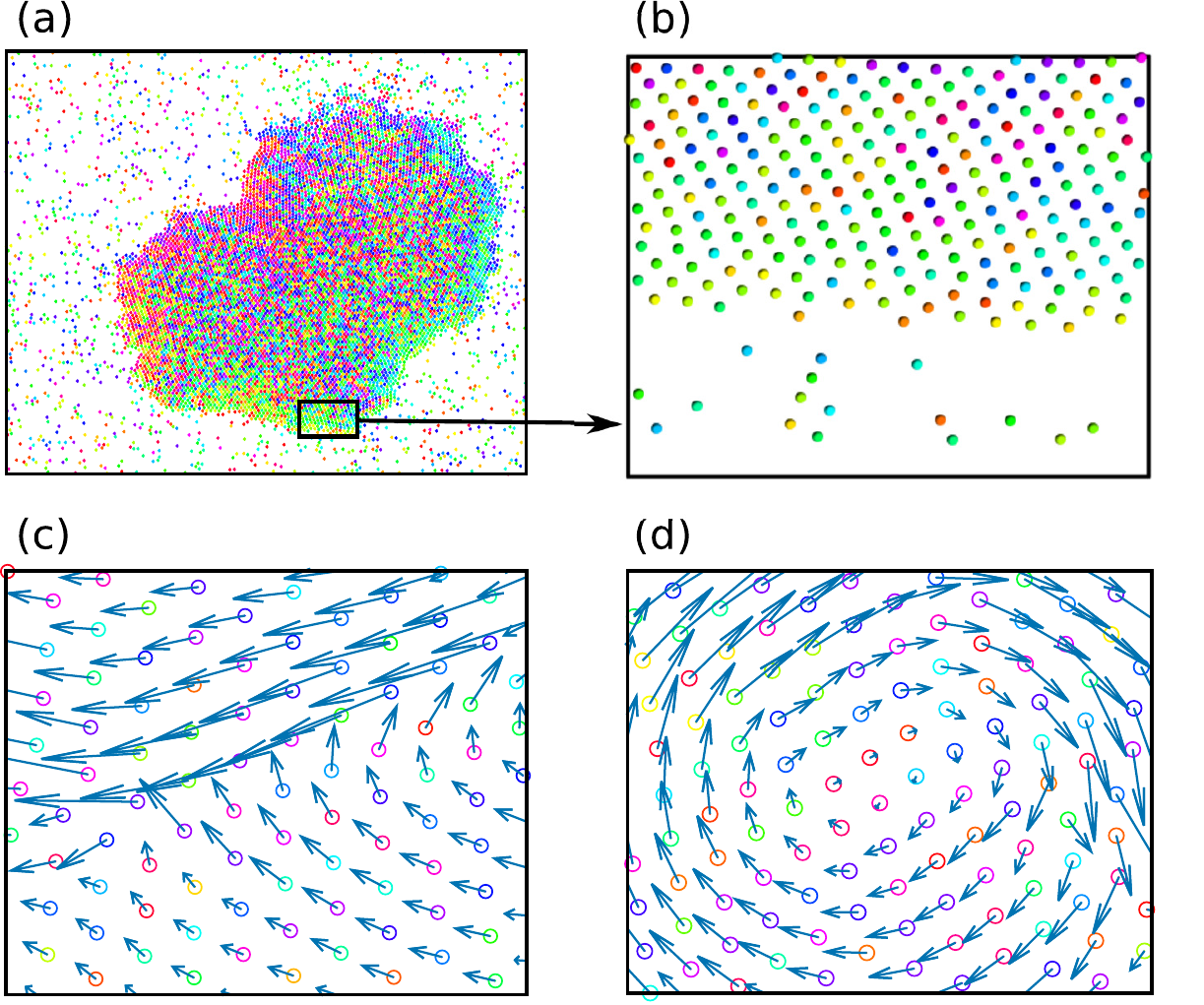}
\caption{In panel (a) we plot a snapshot
  configuration, displaying MIPS, enlarging 
a window near the surface of the cluster.
  Colors encode the self-propulsion direction.
Panel (c) and (d) 
are windows of the bulk where we plot the 
velocities of each particle with blue arrows, showing aligned and
  vortex domains, respectively. Data are obtained by simulation with
  $v_0=50$, $D_r=0.2$ and the other parameters as described in the
  text. }\label{fig:main}
\end{figure}

Numerical integration 
of Eq.~\eqref{eq:xf_dynamics} is performed for a system of $N$ 
particles in a square box of length $L$, with periodic boundary
conditions.  We set a packing fraction $\phi=0.64$, where MIPS is
known to occur at small enough values of
$D_r$~\cite{fily2012athermal}. Indeed, Fig.~\ref{fig:main}(a) shows
the coexistence of a stable dense cluster and a dilute disordered
phase, at $D_r= 0.2$. The boundary of the cluster is highly dynamical:
continuously in time, particles join the cluster and leave it, in such
a way that the average cluster population does not change. In
Fig.~\ref{fig:main} (b-d) we enlarge three representative regions of
the system.  The bulk displays a highly ordered close-packing
configuration~\cite{redner2013structure}.  The study of the pair
correlation function, $g(r)$, shown in the Supplemental Materials (SM), reveals
that the main peak occurs at a distance $\bar{r}<\sigma$ in the
cluster: particles attain a steady-state configuration with large
potential energy, where each microswimmer climbs on the repulsive
potential exerted by the surrounding ones.  Besides, the occurrence of
a second double-split peak reveals a hexagonal lattice structure, in
agreement with the direct observation and previous
studies~\cite{redner2013structure}.
The colors in Figs.~\ref{fig:main}(a-d) encode the orientation,  $\mathbf{n}$, of the
self-propelling force which appears to lack any kind of
alignment.

In Fig.~\ref{fig:main} (c-d) we give evidence of the main novel
phenomenon reported here. We draw with blue arrows the velocities,
$\dot{\mathbf{x}}_i$, of each microswimmer which is in general
different from the orientation of the active force, i.e. $\dot{\mathbf{x}}_i
\neq \mathbf{n}_i v_0$.  
Despite the absence of any alignment interaction,
the velocities of the microswimmers in the bulk of the cluster align,
self-organizing in large oriented domains inside the
cluster.  Even if each $\mathbf{n}_i$ points randomly, particles in large groups move in the same direction (Fig.~\ref{fig:main}~c)).
Such domains dynamically self-arrange continuously in time and, in some cases, evolve into vortex structures as evidenced in Fig.~\ref{fig:main}~d).
The average velocity of each domain is quite smaller than $v_0$ (the
typical speed in the absence of interactions). 
Further details about the velocity distributions in the different phases are contained in the SM.



\begin{figure}[!t]
\centering
\includegraphics[width=0.9\linewidth,keepaspectratio]{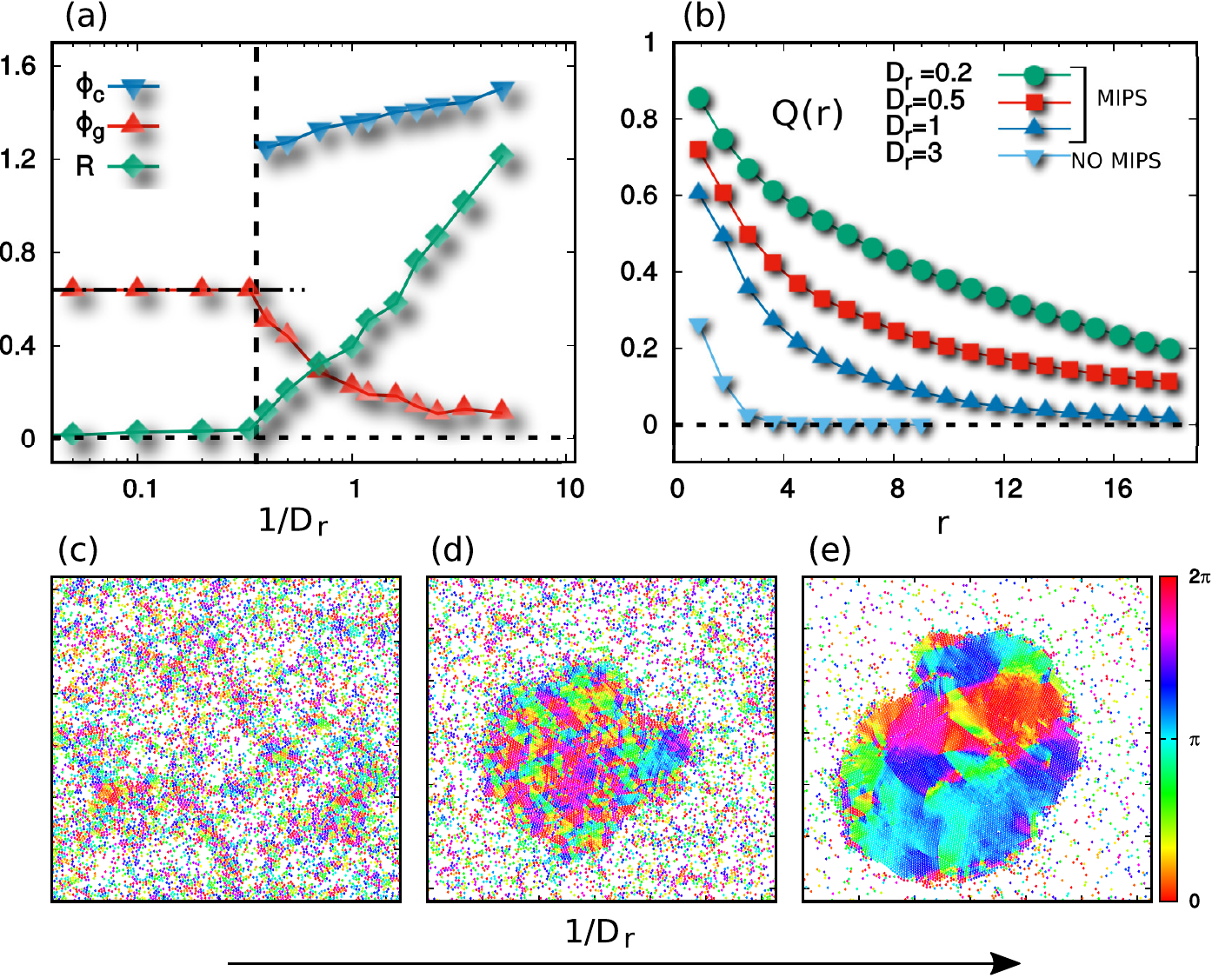}
\caption{ Panel (a): density, $\phi_g$ (red upper triangles) and $\phi_c$ (blue
  lower triangles) for the dilute and the cluster phase, respectively, as a
  function of $1/D_r$. Velocity-alignment order parameter, $R$ (green
  diamonds), as a function of $1/D_r$. For presentation reasons, $R$ is rescaled by a factor 6.
Black dashed lines are
  eye-guides: the vertical one identifies the value of $1/D_r$ at
  which the MIPS-transition occurs. Instead, the horizontal lines refer to
  the nominal density ($\sim 0.64$) and the value of $R$ in absence of
  velocity alignment ($\sim 0$).  Panel (b): $Q(r)$ for different
  values of $D_r$, as shown in the legend, where we specify the
  presence or not of the phase separation.  Panel (c), (d) and (e):
  Snapshot configurations for three different values of $1/D_r$. Panel
  (c) is obtained for $D_r=3$, panel (d) for $D_r=1$ and panel (e) for
  $D_r=0.2$. Colors are associated with the direction of the velocity of
  each particle.  All the simulations are realized with numerical
  density $\sim 0.64$, $v_0=50$ and the other parameters specified in
  the text. }\label{fig:alignment}
\end{figure}

The global alignment of the particles or polarization is commonly measured by considering
 the propulsion orientation, ${\mathbf n}_i$, of each particle,
while here we focus on the velocity $\dot
{\mathbf x}_i$.
A possible order parameter is represented by the sum $\left| \sum_{k=1}^N e^{i \psi_k(t)}\right|$, where $\psi_k$ is the angle formed by the particle velocity
with respect to the $x$ axis.
Such a parameter has the property of being zero for
particles without any alignment while it returns one for perfectly
aligned particles.  Unfortunately, even if restricted to
particles inside a cluster, such a quantity does not reveal a clear polarization of
the system because of the presence of several domains with different
orientations.  
Thus, we introduce the spatial correlation function of the velocity orientation, $Q_i (r)$.
We define the angular distance between two angles $d_{ij}=\min[|\psi_i - \psi_j|, 2\pi - |\psi_i - \psi_j|]$,
and measure the velocity alignment between particle $i$ and the neighboring particles in the circular crown 
of mean radius $r=k\bar{r}$,  with integer $k>0$, and thickness $\bar{r}$, in such a way that 
\begin{equation}
Q_i (r)= 1 - 2\sum_{j}\frac{d_{ij}}{\mathcal{N}_k\pi} 	\,,
\end{equation}
where the sum runs only over the particles in the circular shell
selected by $k$ and $\mathcal{N}_k$ is the number of particles in that
shell.  Then, we define $Q(r)=\sum_i Q_i(r)/N$, which reads 1 for
perfectly aligned particles in the $k$-th shell, $-1$ for anti-aligned
particles and $0$ in  the absence of any form of alignment. $Q(r)$ can
quantify partial alignment even in the absence of global
polarization. Panel (b) of Fig.~\ref{fig:alignment} shows $Q(r)$ for
different values of $D_r$ in a set of simulations with $v_0=50$ (the
other parameters are fixed in the same way as before). In general $Q$
is a decreasing function of $r$. At large $D_r$ where MIPS does not
occur, the alignment measured by $Q(r)$ is absent or very weak, affecting no more than the
first two shells. 
In the MIPS configuration,
the degree of alignment increases
and spans larger and larger distances, when $D_r$ is reduced.
Three snapshots with
color-encoded velocity orientation are shown in panels (c-e) of
Fig.~\ref{fig:alignment}, showing the growth of velocity-aligned
domains in the cluster phase.  In fig.~\ref{fig:alignment}(a) we
investigate the nature of this ordering phenomenon by measuring the
following order parameter
\begin{equation}
R = \int Q(r) dr \,.
\end{equation}
The integral is performed over the whole cluster domain while in the absence of
phase separation we consider the whole box. 

 
To evaluate the relationship between this growing spatial velocity order and
MIPS, we compare $R$ with an established order parameter
for phase separation. Local packing fractions 
show a unimodal distribution when the system is not phase-separated and a
bimodal one when phase separation occurs.  
The height of the peaks in
the distribution identifies the most probable values of the packing fraction 
 in the unimodal case, it corresponds to the homogeneous phase $\phi_g \approx \phi$. Instead, in the bimodal case, the cluster phase is identified by the peak with $\phi_c > \phi$ while the disordered phase by that with $\phi_g < \phi$.
These results are reproduced as a
function of $1/D_r$ in Fig.~\ref{fig:alignment}(a).  At $1/D_r \sim
0.3$ phase separation is revealed by the transition from the single peak
to the double peak in the distribution of the packing fraction. 
In our configuration, $\phi_g$ in the homogeneous phase follows continuously the  values outside the cluster, which forms at a much higher packing fraction.
The comparison with the curve for $R$ reveals the most interesting
information of our study, that is the coincidence between the MIPS
transition and the growing of the velocity-order. Indeed, $R$ reveals a two-steps behavior, 
being almost-zero before $1/D_r \sim 0.3$ and revealing a sharp, monotonic increase starting from this point.


To shed light on the above phenomenology we perform
an exact mapping of the original ABP dynamics, Eqs.~\eqref{eq:wholeABPdynamics}, 
 in the same spirit of the Ornstein-Uhlenbeck (AOUP) model~\cite{fodor2016far, caprini2018linear, caprini2019activity}. 
In particular, we obtain an equation of motion for the microswimmer velocity, $\mathbf{v}_i=\dot{\mathbf{x}}_i$, which is an unprecedented result for ABP.
In two dimensions, $\mathbf{v}_i$ follows: 
\begin{equation}
\label{eq:xv_dynamics}
\mu\dot{\mathbf{v}}_i = - \gamma \sum_{j=1}^N{\Gamma}_{ij}({\mathbf r}_{ij}) \mathbf{v}_j + \mathbf{F}_i + \sqrt{2 \gamma (\mu v_0^2)} \boldsymbol{\xi}_i \times {\mathbf n}_i \,,
\end{equation}  
where $\boldsymbol{\xi}_i$ is the stochastic vector with components
$(0,0, \xi_i)$ and both $\mathbf{v}_i$ and $\mathbf{x}_i$ belong to
the plane $xy$.  The effective mass is $\mu=\gamma/D_r$ and the viscosity matrix $\Gamma_{ij}$ has
the following structure:
\begin{equation}
\label{eq:definitionofGamma}
\Gamma_{ij}^{\alpha \beta}({\mathbf r}_{ij}) = \delta_{ij}\delta_{\alpha \beta} + \frac{1}{D_r \gamma} \nabla_{i\alpha} \nabla_{j \beta} \sum_{k<l} U(|{\mathbf r}_{kl}|) \,,
\end{equation}
where Latin and Greek indices refer to the particle number and the
spatial vector components, respectively.  
The derivation of Eq.~\eqref{eq:xv_dynamics} is reported in the SM.
Eq.~\eqref{eq:xv_dynamics} is
the equation of motion of an underdamped particle under the action of
a space-dependent Stokes force and a multiplicative noise both in the
velocity and in the position of the target microswimmer.  The noise
term always acts perpendicularly to
$\mathbf{n}_i$, because of the cross product. 
The most interesting
information contained in Eq.~\eqref{eq:xv_dynamics} is the fact that
the dynamics of the $i$-th particle is strongly influenced not only by
the positions but also by the velocities of the surrounding particles, through
the matrix $\Gamma_{ij}$ which - because of the factor $1/D_r$ -
is dominated by the velocity coupling terms. 
We recall that
Eq.~\eqref{eq:xv_dynamics} is almost identical to the equation of
motion of interacting AOUP particles~\cite{marconi2016velocity}, the only difference
being the noise term, which in AOUP is additive and uncorrelated,
i.e. $\boldsymbol{\xi}_i \times {\mathbf n}_i$ is replaced by
a noise vector with independent components.

Inside a cluster Eq.~\eqref{eq:xv_dynamics} can be further simplified,
taking advantage of the hexagonal spatial order: we may assume that a
particle in the bulk of the cluster has $6$ neighbors at relative
positions $\bar{\mathbf{r}}_{ij}$ with $j=1..6$, with constant modulus
$\bar{r}=|\bar{\mathbf{r}}_{ij}| <\sigma$, as revealed, for instance, by the $g(r)$.  
With these assumptions, one gets for the particle at the center of the hexagon
\begin{equation}
\label{eq:appox_clusterdynamics}
\mu\dot{\mathbf{v}} = -\frac{1}{D_r}\sum^{6}_{j=1} \hat{H}_{j} \cdot ( \mathbf{v}-\mathbf{v}_j)  - \gamma \mathbf{v}  + \sqrt{2 \gamma (\mu v_0^2)} \boldsymbol{\xi} \times {\mathbf n} \,,
\end{equation}
where $\hat{H}_j$ is the matrix coupling the central particle to the $j$-th particle and its elements depend on $\bar{r}$ and on the angle formed by $\mathbf{x}_{ij}=\mathbf{x}_j - \mathbf{x}_i$ and the $x$-axis.
The matrix elements of $\hat{H}_j$ are reported in the SM.
Equation~\eqref{eq:appox_clusterdynamics} can be rewritten in terms of the average velocity vector of the $6$ neighbors $\mathbf{v}^*=\sum_{j=1}^6 \mathbf{v}_j/6$ and takes the form
\begin{equation}
\label{eq:appox_clusterdynamics2}
\mu\dot{\mathbf{v}} = -\frac{\hat{J}}{D_r} \cdot ( \mathbf{v}-\mathbf{v}^*)  + \frac{1}{D_r}  \sum^{6}_{j=1} \hat{H}_{j}\cdot ( \mathbf{v}_j -\mathbf{v}^*)-\gamma \mathbf{v}  + \mathbf{k} \,,
\end{equation}
with $\hat{J} = \sum_j^6\hat{H}_j=
3\left[U''(\bar{r})+\frac{U'(\bar{r})}{|\bar{r}|}\right] \mathcal{I}$,
being $\mathcal{I}$ the identity matrix and $\mathbf{k}$ the noise vector of Eq.~\eqref{eq:appox_clusterdynamics}. 
 Eqs.~\eqref{eq:appox_clusterdynamics} and \eqref{eq:appox_clusterdynamics2} are derived in the SM. 
We notice that
$\left(U''(\bar{r})+\frac{U'(\bar{r})}{|\bar{r}|}\right)>0$ which
means that the first term in the rhs of
Eq.~\eqref{eq:appox_clusterdynamics2} is a Vicsek-like force aligning
the velocity of the central particle towards the average velocity vector
$\mathbf{v}^*$~\cite{gregoire2004onset}. 
In two special cases the second force in the rhs of
Eq.~\eqref{eq:appox_clusterdynamics2} vanishes: i) trivially when the
$6$ neighbors have identical velocities ${\mathbf v}_j = {\mathbf
  v}^*$; ii) when the $6$ neighbors have velocities arranged according
to a vortex-like pattern. 
This statement is proved in the SM.
In both cases at large $1/D_r$ the
dynamics of $\mu\dot{\mathbf{v}}$ is dominated by the Vicsek-like aligning
force 
 (first term in the rhs of Eq.~\eqref{eq:appox_clusterdynamics2}) and
one has a rapid convergence ${\mathbf v} \to {\mathbf v}^*$.
At the end of this convergence, i.e. when the
velocity of the central particle is exactly aligned with the $6$
neighbors, the aligning force disappears and the sub-dominant bath-like
terms $- \gamma \mathbf{v} + \sqrt{2 \gamma (\mu v_0^2)}
\boldsymbol{\xi} \times {\mathbf n}$ perturb the velocity. 
At this stage, 
the Vicsek-like 
force comes back into
play and restores the alignment. For more general cases (i.e. when the
$6$ neighbors are not aligned or are arranged in a vortex pattern), a second force, depending on the
deviations $\mathbf{v}_j-\mathbf{v}^*$ with a large pre-factor
$1/D_r$, comes into play.
However, when particles are close to alignment,
the terms $\mathbf{v}_j-\mathbf{v}^*$ are small and uncorrelated, 
so that their sum is even smaller and does not alter significantly the
aligning term, as numerically checked. 
A rigorous general estimate of the fate of
Eq.~\eqref{eq:appox_clusterdynamics2} is difficult.

Our analytical description in terms of effective velocities could be adapted to describe the emergent polar order of rod-like~\cite{ginelli2010large, yang2010swarm, bar2019self} or dumbell~\cite{suma2014motility, cugliandolo2017phase} particles, introducing the angular velocity induced by the self-propulsion.

To derive the exponential-like form of the spatial velocities correlations, we assume all particles sitting on an
infinite hexagonal lattice, with each particle's velocity connected to
its $6$ neighbors by Eq.~\eqref{eq:appox_clusterdynamics}. 
Since ${\mathbf n}$ and ${\mathbf v}$ are roughly uncorrelated in the bulk,
we replace the multiplicative noise with an additive uncorrelated noise, as in the AOUP case~\cite{caprini2019active}.
The evolution of this velocity field 
can be mapped, by Fourier transforming, onto a Langevin
equation for each mode in the reciprocal lattice. Its steady-state
solution gives the velocity structure factor or, equivalently, the
spatial correlations of the velocity field. This analysis 
demonstrates that the correlation length of the velocity field reads
\begin{equation}
\label{eq:predictionlambda_main}
\lambda_s \approx \bar{r}\left[ \frac{3}{4\gamma D_r} \left(U''(\bar{r})+\frac{U'(\bar{r})}{|\bar{r}|}\right)\right]^{1/2}  \,,
\end{equation}
whose derivation is reported in the SM. 
This argument suggests a correlation length growing with $1/D_r$
 in qualitative agreement with Fig.~\ref{fig:alignment} a) and b).
We suspect that terms at small wavelengths can be
important, for instance, in the explanation of the vortex structures.


Our study demonstrates an unprecedented strong connection between
 velocity ordering and MIPS transitions. In the absence of any microscopic force
that explicitly aligns velocities, we observe the emergence of
velocity patterns, aligned or vortex-like domains in a dense cluster,
which become more and more pronounced as the persistence of the active
force increases.

We stress here the deep non-equilibrium nature revealed by our study.
Such a velocity order cannot be observed in any passive Brownian suspensions of spherical particles, since,
in those cases, particles' velocities are distributed according to independent Boltzmann distributions.
Thus, the growth of order in the velocity field cannot be explained in
equilibrium-like theories unless an effective aligning force is
introduced in a macroscopic ``Hamiltonian'' which is absent in the
microscopic model.  This would be in line with previous
equilibrium-like approaches where effective attractive interactions
were introduced to explain phase separation~\cite{farage2015effective,
  rein2016applicability} also at the level of an effective free-energy
functional~\cite{tailleur2008statistical, cates2013active, speck2016collective, solon2018generalized} or employing an effective Cahn-Hilliard
equation~\cite{stenhammar2013continuum, speck2014effective}.  All such strategies were already
challenged by observations about pressure~\cite{solon2015pressureNat,
  solon2015pressure}, negative interfacial tension between the
coexisting phases~\cite{bialke2015negative, patch2018curvature} and
different temperatures inside and outside the
cluster~\cite{mandal2019motility}, all inconsistent with any
equilibrium-like scenario.  The phenomenology discussed here represents
an additional argument in favor of a purely non-equilibrium approach.

In virtue of our results, we argue that the full comprehension of MIPS
cannot be obtained in terms of the density field only, but requires,
at least, the employment of another vector field to account for the
velocity alignment.  The introduction of a vectorial field to model
the velocity alignment, for instance in the framework of field
theories~\cite{stenhammar2014phase, wittkowski2014scalar, tjhung2018cluster, grossmann2019particle, solon2018generalizedscalar, paoluzzi2019statistical}, may offer a
new interesting perspective to increase the understanding of MIPS
combined with the alignment phenomenology presented in this
manuscript.





\bibliographystyle{apsrev4-1}

\bibliography{activeMIPS.bib}

\pagebreak
\newpage

\begin{widetext}
\newpage

\section*{Supplemental Material of ``Spontaneous velocity alignment in Motility-induced Phase Separation''}





In this Supplemental Materials, we provide more details about the main phenomenology and the derivations of the analytical results reported in the main text
In Sec.~\ref{Sec1}, we show the pair correlations and the distribution function of the velocity modulus inside and outside the cluster.
Sections~\ref{Sec2} and~\ref{Sec3} are devoted to the detailed derivations of Eq.~(4), Eq.~(6) and Eq.~(7) of the main text, i.e. the equations of motion for the velocity and the effective equation ruling the particles' dynamics inside the cluster.
Instead, the form of the spatial velocity correlation, i.e Eq.~(8) of the main text, is derived in Sec.~\ref{Sec4}.
Finally, In Sec.~\ref{Sec5}, Eq.~(7) is evaluated for the typical velocity-patterns reported in Fig.~1 of the main text, namely aligned and vortex-like domains.


\section{Numerical analysis, pair correlation function and single particle velocity distribution}\label{Sec1}

The numerical analysis of Eqs.(1) of the main text 
has been
performed using a finite-difference scheme with periodic boundary
conditions in a square box of size $L=125$. The number of particles
have been fixed to $N=10^4$, obtaining a packing fraction, $\phi=0.64$.
The WCA potential, described in the main text, is choosen fixing
$\epsilon=1$ and $\sigma=1$, for the sake of simplicity.  We always
fix the self-propulsion strength to $v_0=50$, since we focus on the
effect of the persistence time, $1/D_r$, varied from $10^{-2}$ to
$10$.

To understand the structure of an active suspension of $N$ particles
we study the pair correlation function defined as $g(r) =
\sum_i\sum_{j\neq i} \langle\delta \left(\mathbf{x} -\mathbf{x}_{ij}
\right) \rangle A/N^2$, being $A$ the area occupied by the system, the
sum runs over the distances between the particles' pairs,
$\mathbf{x}_{ij}$ and $\mathbf{x}$ denotes the target distance.  The
brackets indicate a circular average over $\mathbf{x}$ such that
$|\mathbf{x}| = r$.  In Fig.~\ref{fig:FigureSI} a) we evaluate $g(r)$
within (blue curve) and outside (red curve) the cluster for a typical
set of parameters displaying MIPS, namely $v_0=50$ and $D_r=0.2$.  The
pair correlation within the cluster shows the typical solid-like shape
\cite{redner2013structure} with the occurrence of a second split peak,
while $g(r)$ outside the cluster is more similar to the pair
correlation corresponding to a liquid.  The first peak of $g(r)$
inside the cluster, which measures the typical inter-particle distance
between neighboring particles, occurs at a distance $\bar{r}<\sigma$.
This means that particles ``climb on the repulsive potential''.
Instead, $g(r)$ outside the cluster goes rapidly towards one,
displaying only the initial peak, placed at position $\sim \sigma$.
This peak has not a Brownian counterpart, being the density very low:
a Brownian suspension of particles with the same area fraction shows a
peak-less $g(r)$ regardless of the temperature
value~\cite{caprini2019activity}.  The occurrence of such an initial
anomalous peak means that particles prefer to form unstable couples or
small groups at variance with an equilibrium-like gas.

\begin{figure}[!h]
\centering
\includegraphics[width=0.95\linewidth,keepaspectratio]{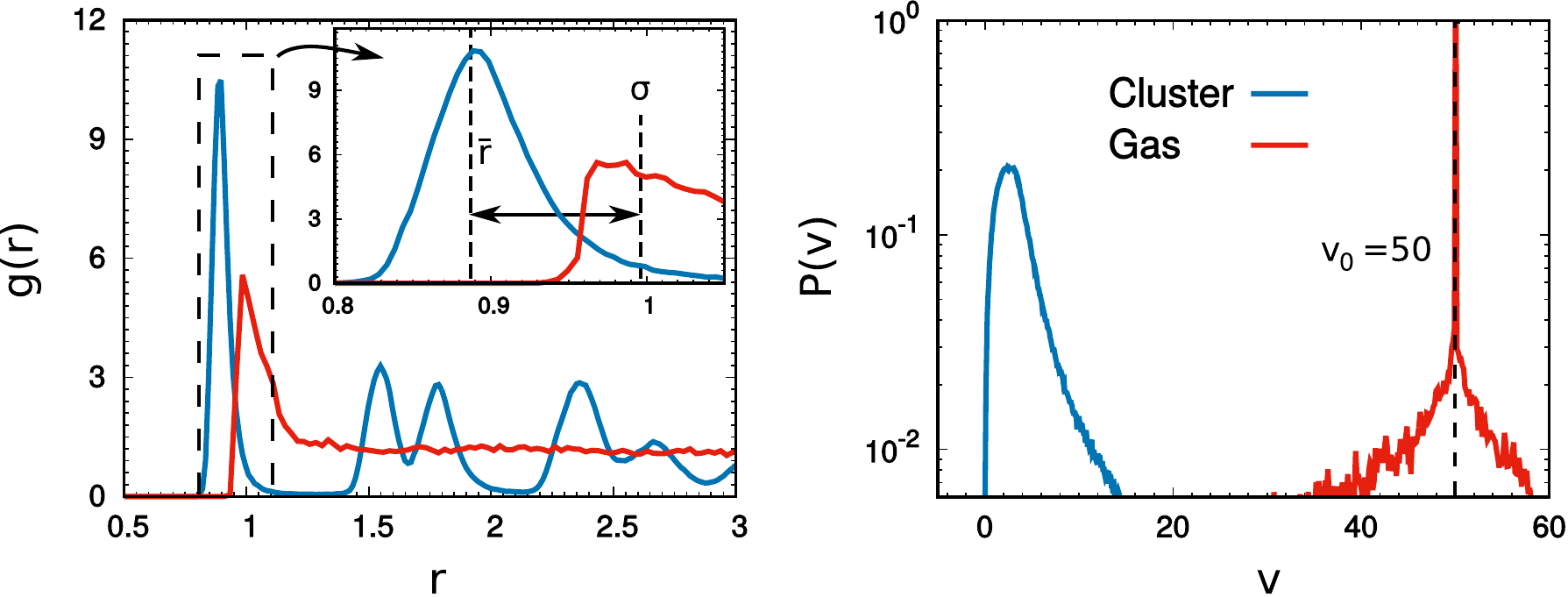}
\caption{ Panel (a): Pair correlation function, $g(r)$, computed within (blue) and outside (red) of the main cluster. The inset enlarges the first peak of the $g(r)$ as shown in the Figure.
Panel (b): probability distribution function of the velocity, $p(v)$, within (blue) and out (red) of the main cluster.
The observables are obtained from a simulation with $v_0=50$ and $D_r=0.2$. 
The other parameters are the same described in the main text.
 }\label{fig:FigureSI}
\end{figure}

In fig.~\ref{fig:FigureSI} b), we also study the probability distribution function, $P(v)$, of the velocity modulus, $v=|\mathbf{v}|$, within (blue) and outside (red) of the cluster for a simulation with $v_0=50$ and $D_r=0.2$, displaying MIPS.
Particles inside the cluster have a mean velocity, $\langle v\rangle$, slower than $v_0$, which is instead the typical speed value of particles in the disordered phase, as emerged by the presence of the large peak at $v=v_0$.
We observe that a consistent fraction of particles in the disordered phase is not interaction-free as revealed by two tails for $v$ smaller and even larger $v_0$.


\section{The velocity of an Active Brownian particle: derivation of Eq.(4)}\label{Sec2}


Eq.~(1b) of the main text, i.e. the dynamics 
of the angle $\theta_i$,
corresponds to the following vectorial equation for the associated orientation vector 
$\mathbf{n}_i$:
\begin{equation}
\label{eq:theta_dynamics1}
\dot{\mathbf{n}}_i= \sqrt{2D_r} \boldsymbol{\xi}_i \times \mathbf{n}_i  \,,
\end{equation}
being $\boldsymbol{\xi}_i$ a three dimensional vector with components $(0,0,\xi_i)$ and $\langle \xi_i(t) \xi_j(t')\rangle=\delta(t-t')$, while $\mathbf{n}_i$ is a unit vector belonging to the $xy$-plane.
In Eq.~\eqref{eq:theta_dynamics1} the noise has multiplicative character and is integrated with the Stratonovich convention.
Taking the time derivative of Eq.~(1a) of the main text 
and defining $\mathbf{v}_i = \dot{\mathbf{x}}_i$, we get:
\begin{equation}
\label{eq:ap_vdynamics}
d\mathbf{v}_i = -\frac{1}{\gamma} \sum_j\nabla_i \nabla_j U_{tot} \cdot \mathbf{v}_j  dt +v_0 d\mathbf{n}_i \,.
\end{equation}
In order to compute the variation $d\mathbf{n}_i $ we switch to Ito calculus and find after some
standard manipulations:
\begin{equation}
\label{eq:ap_dv}
d\mathbf{n}_i = \sqrt{2D_r}\, \boldsymbol{\xi}_i dt \times \mathbf{n}_i - D_r \mathbf{n}_i \,dt \,,
\end{equation}
where by $ \boldsymbol{\xi}_i dt $ we denote the Wiener process $ d\boldsymbol{W}_i= \boldsymbol{\xi}_i dt $.
Putting Eq.~\eqref{eq:ap_dv} into Eq.~\eqref{eq:ap_vdynamics}
we obtain:
\begin{equation}
d\mathbf{v}_i = -\frac{1}{\gamma}\sum_j\nabla_i \nabla_j U_{tot} \cdot \mathbf{v}_j dt -  D_r  v_0\mathbf{n}_i dt   
+v_0\sqrt{2D_r}\,\boldsymbol{\xi}_i dt \times \mathbf{n}_i \nonumber \,.
\end{equation}
Finally, using Eq.~(1a),  
we get:
\begin{equation}
\frac{\gamma}{D_r} d\mathbf{v}_i =-   \gamma \mathbf{v}_i   dt    -\frac{1}{D_r}\sum_j\nabla_i \nabla_j U_{tot} \cdot \mathbf{v}_j dt 
-  \nabla_i U_{tot} dt 
+v_0\sqrt{2 \frac{\gamma^2}{D_r}}\,\boldsymbol{\xi}_i dt \times \mathbf{n}_i \nonumber \,.
\end{equation}
Considering the definition of the matrix $\boldsymbol{\Gamma}$ given by Eq.~(5) 
and $\mu=\gamma/D_r$,  we obtain Eq.~(4) 
of the main text.

\section{Effective equations for particles within the cluster: derivation of Eq.(6) and Eq.(7)}\label{Sec3}

Let us start from Eq.~(4) 
for a system of particles placed on a perfect hexagon, as in the bulk of the cluster. 
A target particle interacts only with its six neighbors at distance $\bar{r}< \sigma$
due to the
the nature of the potential that cuts off the interactions with particles located at distances
larger than $\sigma$.
By symmetry, in Eq.~(4) 
the external force contribution, $\mathbf{F}_i$, on the target particle, 
turns out to be zero  and the only contribution to the dynamics comes from the noise source and from the velocities-dependent terms, $\sum_j\boldsymbol{\Gamma}_{ij} \cdot \mathbf{v}_j$, which explicitly read:
\begin{equation}
\label{eq:app_eq13}
\begin{aligned}
\sum_j\boldsymbol{\Gamma}_{ij} \cdot \mathbf{v}_j&= \sum_j\mathbf{v}_j \cdot \left[\mathcal{I} + \frac{\gamma}{D_r}  \nabla_i \nabla_j  U_{tot} \right] \\
&= \mathbf{v}_i + \frac{\gamma}{D_r} \sum_{j=1}^6\mathbf{v}_i \cdot\nabla_i \nabla_i U \left( r_{ij}  \right) + \frac{\gamma}{D_r}\sum_{j=1}^6 \mathbf{v}_j\cdot  \nabla_i \nabla_j U\left( r_{ij}  \right) \,,
\end{aligned}
\end{equation}
being $r_{ij}$ the distance between the $i$-th and $j$-th particle.
The last two terms of Eq.~\eqref{eq:app_eq13} can be explicitly evaluated by considering the derivative with respect to the spatial components denoted by Greek upper indices:
\begin{equation}
\label{eq:supp_elementH}
 \nabla^{\alpha}_i \nabla^{\beta}_i U\left( r_{ij}  \right) = \left[ U''(r_{ij}) + \frac{U'(r_{ij})}{|r_{ij}|} \right] \frac{r_{ij}^{\alpha}r_{ij}^{\beta}}{|r_{ij}|^2} - \delta_{\alpha\beta}  \frac{U'(r_{ij})}{|r_{ij}|} \,,
\end{equation}
being $r_{ij}^{\alpha} =  r_i^{\alpha} - r_j^{\alpha}$, with $\alpha=x, y$. 
Denoting with $\delta_j$ the angle formed (with respect to the $x$-axis) between the $j$-th and the $i$-th particle, we can note that $r_{ij}^{\alpha}/|r_{ij}|$ reads $\cos{\left(\delta_j\right)}$ and $\sin{\left(\delta_j\right)}$ for $\alpha=x, y$, respectively. 
Since particles belong to a perfect hexagon we can express the angle as a function of $j$ in such a way that $\delta_j = \delta_0 + j\pi/3$. The orientation of the hexagon with respect to the reference frame is fixed by the angle $\delta_0$, which we set to zero for the sake of simplicity.
Expressing the matrix elements of Eq.~\eqref{eq:supp_elementH} in terms of trigonometric functions, we get: 
\begin{equation} \label{pmatrix}
\hat{H}_{j} =  \begin{pmatrix} U''(\bar{r})\cos^2(j \pi/3) +\frac{U'(\bar{r})}{|\bar{r}|}\sin^2(j \pi/3)\;\;\; & \left[ U''(\bar{r}) - \frac{U'(\bar{r})}{|\bar{r}|}\right] \cos(j\pi/3)\sin(j\pi/3) \\ \left[ U''(\bar{r}) - \frac{U'(\bar{r})}{|\bar{r}|}\right] \cos(j\pi/3)\sin(j\pi/3)\;\;\; & U''(\bar{r})\sin^2(j\pi/3) + \frac{U'(\bar{r})}{|\bar{r}|}\cos^2(j\pi/3)  \end{pmatrix}.
\end{equation}
Since the potential depends only on the inter-particle distance the following property holds:
\begin{equation}
\nabla^{\alpha}_i \nabla^{\beta}_j U= - \nabla^{\alpha}_i \nabla^{\beta}_i U \, ,
\end{equation}
and we can easily find Eq.~(6) 
of the main text, assuming that $r_{ij}=\bar{r}$ for every $j$. 

The derivation of Eq.~(7) 
of the main text comes directly from %
 Eq.~(6) ibid., by separating the force $\propto \mathbf{v}$ from the one $\propto \mathbf{v}_j$. 
 In particular, we observe that the sum over $j$ of the matrix element of $\hat{H}_j$ gives rise to a very simple shape in the hexagonal configuration:
\begin{equation}
\sum_{j=1}^6 \hat{H}_j   = 3 \left( U'' +\frac{U'}{\bar{r}}\right) \mathcal{I} \equiv \hat{J}\,.
\end{equation}
Such a simplification comes from the following properties  holding in general for every $\delta_0$:
\begin{flalign}
&\sum_{j=1}^6 \cos^2{\left( \delta_0 + \frac{j \pi}{3} \right)} = \sum_{j=1}^6 \sin^2{\left( \delta_0 + \frac{j \pi}{3} \right)} = 3 \,,\\
&\sum_{j=1}^6 \cos{\left( \delta_0 + \frac{j \pi}{3} \right)} \sin{\left( \delta_0 + \frac{j \pi}{3} \right)} = 0 \, .
\end{flalign}
Finally, adding and subtracting $J \cdot\mathbf{v}^*$, being $\mathbf{v}^*=\sum_{j=1}^6 \mathbf{v}_j$, we obtain Eq.~(7). 

\section{Modes analysis of the velocity field in the hexagonal lattice }\label{Sec4}
In this Section, we derive Eq.~(8) of the main text discussing the approximations involved. 
Let us start from Eq.~(6) 
of the main text:  
Replacing the multiplicative noise term by the additive noise $\sqrt{2 \gamma (\mu v_0^2)} \boldsymbol{\xi}$ and
applying the discrete Fourier transform to the corresponding equation we obtain
\begin{equation}
\mu\frac{\partial }{\partial t}
\tilde{ \mathbf{v}} (\mathbf{k} ,t)
 =-\gamma \tilde {\mathbf{v}}(\mathbf{k} ,t) 
 -\frac{1}{D_r } \tilde H(\mathbf{k} ) \tilde { \mathbf{v}}(\mathbf{k} ,t) 
 +\sqrt{2\gamma \mu v_0^2} \, \tilde{ \boldsymbol{\xi}} (\mathbf{k} ,t) \,,
 \label{dynamicequation51}
\end{equation}
being $\tilde{\mathbf{v}}(\mathbf{k}, t)$ and $\tilde{\boldsymbol{\xi}}(\mathbf{k}, t)$ the Fourier transform of $\mathbf{v}$ and $\boldsymbol{\xi}$, respectively.
The symmetric matrix $ \tilde H(\mathbf{k} ) $, according to Eq.~\eqref{pmatrix}, has the following matrix elements 
\begin{eqnarray}
\tilde H_{xx}(\mathbf{k} )&=&  \left(U''(\bar r)+3 \frac{U'(\bar r)}{\bar r}\right)\left[\cos( \frac{k_x \bar r}{2}) \cos(\frac{ \sqrt 3 k_y \bar r}{2} )-1\right]  +2  \, U''(\bar r)  [\cos(  k_x \bar r )-1]   \,,
\\
\tilde H_{yy}(\mathbf{k} )&=&   \left(3 U''(\bar r)+ \frac{U'(\bar r)}{\bar r} \right) \left[\cos( \frac{k_x \bar r}{2}) \cos(\frac{ \sqrt 3 k_y \bar r}{2} )-1\right]  +2  \frac{U'(\bar r)}{\bar r}  \,   [\cos(  k_x \bar r )-1]   \,,
\\
\tilde H_{xy}(\mathbf{k} )&=& \sqrt{3} \left(U''(\bar r)+3 \frac{U'(\bar r)}{\bar r} \right)  \sin(  \frac{k_x \bar r}{2}) \sin( \frac{ \sqrt 3 k_y \bar r}{2} )  \,.
\end{eqnarray}
Eq.~\eqref{dynamicequation51} can be easily solved
\begin{equation}
\tilde{ \mathbf{v}} (\mathbf{k} ,t)=\tilde{ \mathbf{v}} (\mathbf{k} ,0) e^{-\alpha(\mathbf{k} )t}+\sqrt{2\gamma \mu v_0^2} \, 
\int_0^t  dt'\,e^{-\alpha(\mathbf{k} )(t-t')}  \, \tilde{ \boldsymbol{\xi}}(\mathbf{k} ,t) \,,
\end{equation}
where, for the sake of simplicity, we report $\alpha(\mathbf{k})$ in the small $k$ limit, obtaining:
\begin{equation}
\alpha(\mathbf{k})=D_r+ \frac{3 }{4 \gamma} \left( U''(\bar r)+ \frac{U'(\bar r)}{\bar r} \right)    \left|\mathbf{k}\right|^2  \bar r^2 \,.
\end{equation}
The corresponding equal time velocity-correlation is
\begin{equation}
\label{eq:vvcorr_kspace}
\langle \hat v_x(\mathbf{k},t) \hat v_x(-\mathbf{k},t) \rangle +\langle \hat v_y(\mathbf{k},t) \hat v_y(-\mathbf{k},t) \rangle  = \frac{2 v_0^2}{1 +\lambda_s^2   \left|\mathbf{k}\right|^2  } \,,
\end{equation}
where
\begin{equation}
\label{eq:lambda_prediction_appendix}
\lambda_s \approx \bar{r}\left[ \frac{3 }{4}  \frac{1}{\gamma D_r} \left(U''(\bar{r})+\frac{U'(\bar{r})}{|\bar{r}|}\right)\right]^{1/2} \,.
\end{equation}
The expression~\eqref{eq:lambda_prediction_appendix} corresponds to Eq.~(8)
of the main text. 
Coming back to the real space representation, Eq.~\eqref{eq:vvcorr_kspace} turns into: 
\begin{equation}
\label{eq:spacecorrelation_approx}
\langle   \mathbf{v}( \mathbf{x+r},t)   \mathbf{v}(\mathbf{x},t) \rangle\approx  2 v_0^2 \Bigl(\frac{\lambda_s}{8\pi r} \Bigr)^{1/2} e^{-  r/\lambda_s} \,.
\end{equation}
We outline that the correlation length, Eq.~\eqref{eq:lambda_prediction_appendix}, and the exponential shape of the space correlation, Eq.~\eqref{eq:spacecorrelation_approx}, are the results of the expansion for small $\mathbf{k}$.

\section{Forces contributions in the aligned and vortex domains}\label{Sec5}

In this Section, we calculate the velocity dependent force on a target particle due to the six surrounding particles having velocities, $\mathbf{v}_j$, with $j=1, ... , 6$.
The particle with $j=1$ is placed on the $x$ direction at coordinates $(\bar{r}, 0)$.
The others are placed sequentially in the anti-clockwise sense at reciprocal angular distance $\pi/3$ and at distance $\bar{r}$ from the origin of the reference frame.
We check that in the ideal cases of aligned domains and vortex structures the only relevant force contribution in Eq.~(7) is the alignment term, $ \propto\hat{J} \cdot ( \mathbf{v}-\mathbf{v}^*)$, while the other forces vanish or are irrelevant. 
Let us start from Eq.~(7) of the main text, which we rewrite below, for completeness: 
\begin{equation}
\label{eq:appox_clusterdynamics2bis}
\mu\dot{\mathbf{v}} = -\frac{1}{D_r} \hat{J} \cdot ( \mathbf{v}-\mathbf{v}^*)  + \frac{1}{D_r}  \sum^{6}_{j=1}( \hat{H}_{j} -\frac{\hat{J}}{6})\cdot  \mathbf{v}_j -\gamma \mathbf{v}  + \sqrt{2 \gamma (\mu v_0^2)} \boldsymbol{\xi} \times {\mathbf n} \,,
\end{equation}
The last two terms of the right-hand side of Eq.~\eqref{eq:appox_clusterdynamics2bis} are irrelevant in the large persistence regime, where $D_r$ is small.
Instead, the second addend of the right-hand side of Eq.~\eqref{eq:appox_clusterdynamics2bis} needs to be computed:
\begin{equation}
{\bf T}\equiv \frac{1}{D_r}\sum^{6}_{j=1}( \hat{H}_{j} -\frac{\hat{J}}{6})\cdot  \mathbf{v}_j  \,.
\end{equation}
By symmetry, the contributions on {\bf T} due to the particles placed at he opposite vertices of the hexagon are equal.
Thus, in our notation, we have $H_1=H_4$, $H_2=H_5$ and $H_3=H_6$.
Below, we write explicitly each term:
\begin{flalign}
&\hat{H}_{1}-\frac{\hat{J}}{6} =\hat{H}_{4}-\frac{\hat{J}}{6}= \left(U''(\bar{r}) -\frac{U'(\bar{r})}{\bar{r}}\right) \begin{pmatrix} -\frac{1}{4}\;\;\; &  \frac{\sqrt{3}}{4} \\ \frac{\sqrt{3}}{4} \;\;\; & \frac{1}{4}   \end{pmatrix} \,,\\
&\hat{H}_{2}-\frac{\hat{J}}{6} =\hat{H}_{5}-\frac{\hat{J}}{6}= \left(U''(\bar{r}) -\frac{U'(\bar{r})}{\bar{r}}\right) \begin{pmatrix} -\frac{1}{4}\;\;\; &  -\frac{\sqrt{3}}{4} \\ -\frac{\sqrt{3}}{4} \;\;\; & \frac{1}{4}   \end{pmatrix} \,,\\
&\hat{H}_{3}-\frac{\hat{J}}{6} =\hat{H}_{6}-\frac{\hat{J}}{6}= \left(U''(\bar{r}) -\frac{U'(\bar{r})}{\bar{r}}\right) \begin{pmatrix} \frac{1}{2}\;\;\; &  0 \\ 0 \;\;\; & -\frac{1}{2}\end{pmatrix} \,.
\end{flalign}
Using the above expressions for $H_j$ we get:
\begin{flalign}
T_x&=\frac{1}{4 D_r} \left(U''(\bar{r}) -\frac{U'(\bar{r})}{\bar{r}}\right) \left[2 v_{6x}+2 v_{3x}-v_{1x}-v_{2x}
-v_{4x}-v_{5x}+\sqrt 3 (v_{1y}+v_{4y}-v_{2y}-v_{5y})     \right] \,,\\
T_y&=\frac{1}{4 D_r} \left(\frac{U'(\bar{r}) - U''(\bar{r}) }{\bar{r}}\right) \left[2 v_{6y}+2 v_{3y}-v_{1y}-v_{2y}
-v_{4y}-v_{5y}+\sqrt 3 (v_{1x}+v_{4x}-v_{2x}-v_{5x})     \right] \,.
\end{flalign}
Both components of the force vanish in the following cases: i) when all velocities are identical, i.e. in the case of aligned domains.
ii) When the velocity of the six neighboring particles are arranged in a vortex configuration, for instance, described by the following velocity profile:
\begin{equation}
{\mathbf{v}}_j=v_0 \left[ -\sin\left(j\frac{\pi}{3}\right) , \cos\left(j\frac{\pi}{3}\right) \right] \,.
\end{equation}
In this last case, the corresponding average velocity vaninshes, i.e. $\mathbf{v}^*=0$.

\end{widetext}

\end{document}